\documentclass[3p,twocolumn,times]{elsarticle}

\usepackage{hyperref}

\journal{Journal of \LaTeX\ Templates}

\usepackage{amsmath, amssymb, amsfonts}
\usepackage{graphicx}

\newtheorem{theorem}{Theorem}

\newcommand{\tens}{\mathsf}
\newcommand{\NN}{\mathbb{N}}

\renewcommand{\ge}{\geqslant}

\begin{document}

\begin{frontmatter}

\title{Enumerating secondary structures and structural moieties for circular
RNAs}

\author[add1,add2,add3,add4]{Jose A. Cuesta\corref{ca}}
\cortext[ca]{Corresponding author}
\ead{cuesta@math.uc3m.es}

\author[add1,add5]{Susanna Manrubia}
\ead{smanrubia@cnb.csic.es}

\address[add1]{Grupo Interdisciplinar de Sistemas Complejos (GISC)}
\address[add2]{Departamento de Matem\'aticas, Universidad Carlos III de Madrid,
Spain}
\address[add3]{Institute for Biocomputation and Physics of Complex Systems,
Zaragoza, Spain}
\address[add4]{UC3M-BS Institute of Financial Big Data (IFiBiD)}
\address[add5]{National Biotechnology Centre (CSIC), Madrid, Spain}

\begin{abstract}
A quantitative characterization of the relationship between molecular sequence
and structure is essential to improve our understanding of how function
emerges. This particular genotype-phenotype map has been often studied in the
context of RNA sequences, with the folded configurations standing as a proxy
for the phenotype. Here, we count the secondary structures of circular RNAs of
length $n$ and calculate the asymptotic distributions of different structural
moieties, such as stems or hairpin loops, by means of symbolic combinatorics.
Circular RNAs differ in essential ways from their linear counterparts. From the
mathematical viewpoint, the enumeration of the corresponding secondary
structures demands the use of combinatorial techniques additional to those used
for linear RNAs. The asymptotic number of secondary structures for circular
RNAs grows as $a^n n^{-5/2}$, with $a$ depending on particular constraints
applied to the secondary structure. As it occurs with linear RNA, the abundance
of any structural moiety is normally distributed in the limit $n \to \infty$,
with a mean and a variance that increase linearly with $n$.
\end{abstract}


\end{frontmatter}


\section{Introduction}

\label{intro}
Notwithstanding the important role that selection has traditionally played in
evolutionary theory, evolution is not possible if selection has not variation
to act upon. Thus mutations ---widely understood as imperfect replications---
are the fuel to evolutionary dynamics. But mutations act at the level of the
\emph{genotype} whereas selection acts at the level of the \emph{phenotype}
---the physical manifestation of the genotype---, and the translation from one
to the other ---the so-called genotype-phenotype (GP) map--- is far from
trivial~\cite{wagner:2011c}.  Most mutations have no effect on the phenotype 
(they are neutral), whereas occasionally a mutation has a dramatic (mostly 
deleterious but sometimes beneficial) phenotypic effect. Thus, evolutionary 
dynamics is critically affected by the structure of the GP 
map~\cite{alberch:1991}.

Understanding the GP map is a challenge for the evolutionary community, overall
because addressing this problem in real systems is of an overwhelming 
complexity. Accordingly, several simplified models have been studied to gain 
insights into this difficult issue~\cite{wagner:2011}. Computationally 
tractable models incorporate only a few levels among those involved in an actual
GP map. They have considered protein folding~\cite{dill:1985,li:1996} or 
protein aggregation~\cite{ahnert:2010} at basic molecular levels, and 
gene-regulatory~\cite{ciliberti:2007} or metabolic~\cite{matias:2009} networks
at higher functional levels. Recent models encompass different levels at the 
same time~\cite{arias:2014}: In contrast with simple sequence-structure GP
maps, the inclusion of different levels from genotype to phenotype permits
the emergence of properties such as environment-dependent molecular function. 

Pioneer among those models was the folding of sequences of RNA into their
secondary structure ---taken as a proxy for
function~\cite{fontana:1993,schuster:1994}, which likely represents the most
studied GP map to date. Folding is driven by base pair stacking mainly and also
by the formation of hydrogen bonds between {\tt CG}, {\tt AU}, and {\tt GU}
base pairs, and the secondary structure of the molecule is determined by its
minimum free-energy configuration. Despite its apparent simplicity and the
inherent impossibility to capture all features of natural GP relationships, RNA
sequence-to-secondary structure maps have properties shared by all GP maps
studied to date, as the relationship between the number of genotypes yielding
the same phenotype and the neutrality of the
latter~\cite{aguirre:2011,greenbury:2015}. 

An important question in characterizing this GP map is how many different 
secondary structures an RNA molecule $n$ base pairs long can form. 
That problem was solved long ago, with the help of recurrence equations and 
subsequent generating functions, for several variants of the model
\cite{waterman:1978a,waterman:1978b,howell:1980}. Asymptotic expressions
were provided when $n$ is large under different constraints imposed to  
the secondary structure ---such as having a minimum number of unpaired
nucleotides in hairpin loops or stems of a minimal given length. Another
relevant question, which represents a step forward in the relation between
structure and function, is how many secondary structures present particular
structural moieties~\cite{hofacker:1998,nebel:2002}. A prominent example is that of 
short sequences with hairpin loops, which have been shown to act as ribozymes with
ligase catalytic activity under general conditions~\cite{fedor:1999}. This
undemanding phenotype-to-function map could have been essential in the
emergence of RNA molecules with complex activity in a prebiotic RNA
world~\cite{briones:2009}. Beyond characterizing the GP map, having closed-form
expressions for the number of RNA structures with specific structural moieties
is important when comparing structure formation by natural sequences with that
of shuffled versions of the same sequence
\cite{seligmann:2016a,seligmann:2016b}.

The distribution of the number of different structural motifs (stems and hairpin 
loops among others) in the limit of $n$ large has been shown to converge to a 
Gaussian in the limit of large $n$~\cite{reidys:2011,poznanovic:2014}. Two 
different techniques 
employed to reach that goal are symbolic methods introduced in modern 
combinatorics~\cite{flajolet:2009}, as in~\cite{reidys:2011}, and Knudsen-Hein
stochastic context-free grammars~\cite{knudsen:2003}, as 
in~\cite{poznanovic:2014}. In an exhaustive work~\cite{reidys:2011}, Reidys tackled 
in depth the properties of RNA folded structures bearing a type of tertiary 
interactions known as pseudoknots. The functional form of the number of structures 
with pseudoknots as a function of sequence length $n$ is of the general form 
$a^n n^{-b}$, with $a \in \mathbb R^+$ and $b \in \mathbb Q^+$ ---their values 
depending on restrictions put on the folded structure. An important constraint
is the complexity of pseudoknots, which conditions the mathematical description
of the problem. Specifically, folded RNA molecules are first reduced to a core
skeleton containing information only on the pseudoknot architecture of the 
fold. Generating functions for the number of possible alternative core 
structures with the previous architecture are derived and, subsequently, full
folds are recovered by reintroducing stems and unpaired nucleotides in all 
possible compatible positions ---through composition of suitably defined 
generating functions. Eventually, the total number of structures with the 
required pseudoknot properties and other possible structural constraints is
obtained. Further details can be found in~\cite{reidys:2011}. This tricky 
procedure for structures with pseudoknots is not necessary in the case of plain 
secondary structures, as we show here. Application of symbolic combinatorics to 
the latter case serves as an introduction to the calculation of the number of 
secondary structures for circular RNA sequences. As will be shown, particular 
properties of circular RNA demand the introduction of combinatorial techniques 
beyond those needed to enumerate open RNA sequences ---with or without 
pseudoknots. 

Circular RNAs form covalently closed continuous loops with specific properties
that distinguish them from linear RNAs. Among others, circular RNAs are small
and non-coding in most cases, and have higher resistance to
exonuclease-mediated degradation and higher structural stability. Viroids,
first described half a century ago~\cite{diener:1967}, are a relevant example
of circular RNA. These pathogenic, naked RNA molecules of a few hundred
nucleotides in length infect plants, occasionally causing strong symptoms. The
mechanisms implied in cell entry, replication and propagation are still partly
unknown. Viroids present secondary structures with highly conserved regions
that fall within two structural classes: rod-like and branched folds. The
secondary structure of viroids plays an essential role in chemical
function~\cite{flores:2012} and acts as a buffer to control the structural
effect of point mutations~\cite{manrubia:2013}. Virusoids are another class of
circular RNAs that depend on helper viruses for replication and encapsidation.
They are related to viroids, though virusoids code for some proteins. Two
interesting examples in this class of hyperpathogens are Hepatitis delta virus
\cite{saldanha:1990} and the smallest known circular RNA in the viroid-virusoid
class, with 220nt \cite{abouhaidar:2014}. As in viroids, the secondary
structure of virusoids is highly compact and constrained by function.  Circular
RNAs encoded in animal genomes, on the other hand, are currently a hot
topic~\cite{memczak:2013}. Indeed, recent studies report a previously
unsuspected abundance of circular RNAs, which awakes the hunch that they must
play main functional roles in the cell~\cite{salzman:2016}. While some of those
circular RNAs have gene regulatory activity, the function performed by thousand
of others is as yet unknown~\cite{memczak:2013,jeck:2014}. Therefore, a
theoretical understanding of the structural diversity of secondary structures
of circular RNAs appears as a timely endeavor, further considering that closed
RNA sequences have folding restrictions different from those of their linear
counterparts. Formal studies on the folding properties of circular RNAs are
limited, to the best of our knowledge, to the case of symmetric
sequences~\cite{hofacker:2012}, whose contribution to the total number of
sequences and folds asymptotically vanishes as $n$ grows. As we demonstrate
here, specific properties of circular RNA entail a comparatively lower number
of secondary structures and lead to different asymptotic behavior.

The paper is organized as follows. Section~\ref{sec:brief} briefly introduces
those aspects of the symbolic method~\cite{flajolet:2009} relevant for our
study. In Section~\ref{sec:RNA} we derive the generating function for the
number of secondary structures with stems of length at least $s$ and hairpins
with at least $m$ unpaired nucleotides, and recover the known expressions in
the limit $n \to \infty$. Section~\ref{sec:basepairs} contains the calculation
of the frequency of structures with a given number of base pairs and is
followed by the simultaneous count of the number of hairpins in
Section~\ref{sec:2elements}. The method extends to multivariate analysis
suitable for counting combinatorial structures with any number of constraints,
in agreement with results obtained in~\cite{poznanovic:2014}. Though these 
sections mostly review results that in one or another form can be found in the
mathematics literature, we believe it is convenient to rephrase certain aspects 
that are later used, in order to convey a biological intuition of how calculations 
are performed and to make this work self-contained. 
Section~\ref{sec:rings} introduces the main novelty of this work, that is, the
enumeration of secondary structures in circular RNAs, followed by a derivation
of the distributions of base pairs and hairpins as a function of $n$ in
Section~\ref{sec:elementsRings}. We close with a brief discussion. 

\section{Methods}
\label{sec:brief}

A full account of symbolic methods in combinatorics can be found in Part A of
Ref.~\cite{flajolet:2009}. We provide a very brief account in this section. Readers 
familiar with this method can safely skip this section.

A combinatorial class $\mathcal{A}$ will be a set of elements on which a
\emph{size} function $|\cdot|$ is defined. The counting problem is to obtain
$a_n$, the number of elements $a\in\mathcal{A}$ such that $|a|=n$. A related
problem is to obtain the generating function
\begin{equation}
A(z)=\sum_na_nz^n=\sum_{a\in\mathcal{A}}z^{|a|}
\end{equation}
($n$ runs on all possible sizes) whose coefficients yield the sequence
$\{a_n\}$. The second writing for $A(z)$ turns out to be very useful when
thinking about these problems, because it means that every element of
$\mathcal{A}$ contributes to the sum defining $A(z)$ with as many factors $z$
as its size.

If a second function is defined on the elements of $\mathcal{A}$, namely
$\varphi(a)=l$ (representing any other feature of $a$), we can introduce the
bivariate generating function
\begin{equation}
A(z,u)=\sum_n\sum_la_{n,l}z^nu^l=\sum_{a\in\mathcal{A}}z^{|a|}u^{\varphi(a)}.
\end{equation}
Clearly $a_{n,l}$ counts the number of elements in $\mathcal{A}$ of size $n$
and feature value $l$, and the second writing can be interpreted as every
element $a\in\mathcal{A}$ adding to the generating function ---besides the
factor $z^n$--- as many factors $u$ as the value of the feature.

We can combine combinatorial classes to obtain new combinatorial classes. We
first have the combinatorial product $\mathcal{C}=\mathcal{A}\times\mathcal{B}$,
which is the set made of the `composite objects' $ab$, where $a\in\mathcal{A}$
and $b\in\mathcal{B}$ (notice that $ab$ and $ba$ are in general different
objects). The size of the set $\mathcal{C}$ is defined as $|ab|=|a|+|b|$ (the
size of the composite object is the sum of the sizes of the components).
Accordingly,
\begin{equation}
C(z)=\sum_{c\in\mathcal{C}}z^{|c|}=\sum_{a\in\mathcal{A}}\sum_{b\in\mathcal{B}}
z^{|ab|}=\sum_{a\in\mathcal{A}}\sum_{b\in\mathcal{B}}z^{|a|+|b|}=A(z)B(z).
\end{equation}

Another operation is the combinatorial sum, $\mathcal{C}=\mathcal{A}+
\mathcal{B}$, also referred to as disjoint union. $\mathcal{C}$ is the union of
$\mathcal{A}$ and $\mathcal{B}$ provided the elements of these two sets are
distinguishable (in other words, it is as if we paint the elements of these two
sets with two different colors and then make the union of them both). Therefore
$c\in\mathcal{C}$ is either an element of $\mathcal{A}$ or an element of
$\mathcal{B}$ and inherits the corresponding size. Hence,
\begin{equation}
C(z)=\sum_{c\in\mathcal{C}}z^{|c|}=\sum_{a\in\mathcal{A}}z^{|a|}+
\sum_{b\in\mathcal{B}}z^{|b|}=A(z)+B(z).
\end{equation}

There are further more complex operations with combinatorial classes. Thus
\begin{equation}
\mathcal{C}=\mathsf{SEQ}(\mathcal{A}):=\mathcal{E}+\mathcal{A}
+\mathcal{A}\times\mathcal{A}+\mathcal{A}\times\mathcal{A}\times\mathcal{A}
+\cdots,
\end{equation}
where $\mathcal{E}=\{\varepsilon\}$, the class made of the \emph{null} element
alone ($|\varepsilon|=0$), is referred to as the \emph{sequence} of
$\mathcal{A}$, i.e., the combinatorial class made of the null element, plus all
elements of $\mathcal{A}$, plus all pairs of elements of $\mathcal{A}$, and so
on. By applying the transformation rules for the sum and the product
\begin{equation}
C(z)=1+A(z)+A(z)^2+A(z)^3+\cdots=\frac{1}{1-A(z)}.
\end{equation}

Sequences can be constrained to have composite elements just of certain
specific compositions. For instance, $\mathsf{SEQ}_k(\mathcal{A}):=
\mathcal{A}\times\mathcal{A}\times\cdots\times\mathcal{A}$ ($k$ times) is
restricted to sequences made of exactly $k$ elements of $\mathcal{A}$ ---its
generating function being $A(z)^k$. Likewise
\begin{equation}
\begin{split}
\mathcal{C}_{\ge k} &=\mathsf{SEQ}_{\ge k}(\mathcal{A})=\sum_{j=k}^{\infty}
\mathsf{SEQ}_k(\mathcal{A}), \\
\mathcal{C}_{<k} &=
\mathsf{SEQ}_{<k}(\mathcal{A})=\sum_{j=0}^{k-1}\mathsf{SEQ}_k(\mathcal{A}),
\end{split}
\end{equation}
define sequences containing at least $k$ and less than $k$ elements of
$\mathcal{A}$ respectively. Then
\begin{equation}
\begin{split}
C_{\ge k}(z) &=\frac{A(z)^k}{1-A(z)}, \\
C_{<k}(z) &=1+A(z)+A(z)^2+\cdots+A(z)^{k-1} \\
&=\frac{1-A(z)^k}{1-A(z)},
\end{split}
\end{equation}
are their corresponding generating functions.

Other interesting operations with combinatorial classes are power sets
($\mathsf{PSET}$), multisets ($\mathsf{MSET}$), and cycles ($\mathsf{CYC}$)
\cite{flajolet:2009}.

$\mathsf{PSET}(\mathcal{A})$ is the class whose members are made of subsets of
elements of $\mathcal{A}$. Thus
\begin{equation}
\mathcal{C}=\mathsf{PSET}(\mathcal{A}):=\prod_{a\in\mathcal{A}}
\big(\mathcal{E}+\{a\}\big)
\end{equation}
and therefore
\begin{equation}
\begin{split}
C(z) &=\prod_{a\in\mathcal{A}}\left(1+z^{|a|}\right)
=\prod_{n=1}^{\infty}\left(1+z^n\right)^{a_n} \\
&=\exp\left\{\sum_{k=1}^{\infty}\frac{(-1)^{k+1}}{k}A\big(z^k\big)\right\}.
\end{split}
\end{equation}
(The last step follows by writing the product as the exponential of a sum of
logarithms and then Taylor-expanding those logarithms.)

$\mathsf{MSET}(\mathcal{A})$ is the class whose members are made of sequences
of arbitrary length of elements of $\mathcal{A}$. Thus
\begin{equation}
\mathcal{C}=\mathsf{MSET}(\mathcal{A}):=\prod_{a\in\mathcal{A}}
\mathsf{SEQ}\big(\{a\}\big)
\end{equation}
and therefore
\begin{equation}
\begin{split}
C(z) &=\prod_{a\in\mathcal{A}}\left(1-z^{|a|}\right)^{-1}
=\prod_{n=1}^{\infty}\left(1-z^n\right)^{-a_n} \\
&=\exp\left\{\sum_{k=1}^{\infty}\frac{1}{k}A\big(z^k\big)\right\}.
\end{split}
\end{equation}

$\mathsf{CYC}(\mathcal{A})$ is the class whose members are made of circular
sequences of arbitrary length of elements of $\mathcal{A}$. The derivation of
the generating function of $\mathcal{C}=\mathsf{CYC}(\mathcal{A})$ is more
involved \cite[\S A.4]{flajolet:2009}, but can be written in terms of Euler's
totient function $\varphi(k)$ as\footnote{$\varphi(1)=1$, and
$\varphi(k)=p_1^{n_1-1}(p_1-1)\cdots p_r^{n_r-1}(p_r-1)$ if $k=p_1^{n_1}\cdots
p_r^{n_r}$ is the prime factorization of $k>1$. Thus $\varphi(2)=1$,
$\varphi(3)=2$, $\varphi(4)=2$, $\varphi(5)=4$, etc.}
\begin{equation}
C(z)=-\sum_{k=1}^{\infty}\frac{\varphi(k)}{k}\log\left[1-A\big(z^k\big)\right].
\end{equation}

One last class we will need is $\mathsf{MSET}_2(\mathcal{A})=\mathsf{CYC}_2
(\mathcal{A})$, whose members are pairs of elements of $\mathcal{A}$ regardless
of the order (when the order matters the class is $\mathcal{A}\times
\mathcal{A}$).  There are many ways to obtain its corresponding generating
function, but perhaps the easiest is to first introduce
$\mathsf{DIAG}(\mathcal{A})$, the class of pairs of identical elements of
$\mathcal{A}$. Its corresponding generating function is $A\big(z^2\big)$
---because it contains one element per element of $\mathcal{A}$, but its size
is double. Then, $\mathcal{C}=
\mathsf{CYC}_2(\mathcal{A}):=\frac{1}{2}\big[\mathcal{A}\times\mathcal{A}+
\mathsf{DIAG}(\mathcal{A})\big]$, and its generating function will be
\begin{equation}
C(z)=\frac{1}{2}\big[A(z)^2+A\big(z^2\big)\big].
\end{equation}
Further classes and development can be found in \cite{flajolet:2009}.

By way of illustration, consider the class $\mathcal{T}$ of all binary trees
with $n$ interior nodes. This class contains the tree with no interior nodes
$\mathcal{E}$ plus all trees made of a root node $\mathcal{U}=\{\bullet\}$ from
which two new trees of $\mathcal{T}$ hang. Thus
\begin{equation}
\mathcal{T}=\mathcal{E}+\mathcal{T}\times\mathcal{U}\times\mathcal{T}.
\label{eq:binarytrees}
\end{equation}
The size of the tree in $\mathcal{E}$ is zero, whereas the root node
$\mathcal{U}$ ---obviously interior--- contributes $z$ to $T(z)$. Therefore
\eqref{eq:binarytrees} translates into $T(z)=1+zT(z)^2$, whence
\begin{equation}
T(z)=\frac{1-\sqrt{1-4z}}{2z}=\sum_{n=0}^{\infty}\frac{1}{n+1}
\binom{2n}{n}z^n,
\end{equation}
the generating function of Catalan's numbers. A nice property of generating
functions is that we do not need to know the coefficients to obtain their
asymptotic expression. For that we can resort to an extension of Darboux's
theorem \cite{flajolet:2009,hofacker:1998}:

\begin{theorem}[Darboux]
Let $f(z)=\sum\limits_{n=0}^{\infty}f_nz^n$, with $f_n\ge 0$, be an analytic
function in the circle $|z|<\zeta$ of the form
\begin{equation}
f(z)=g(z)+h(z)\left(1-\frac{z}{\zeta}\right)^{\alpha}
+O\left(\left(1-\frac{z}{\zeta}\right)^{\alpha+1}\right), \quad
\alpha\notin\NN,
\end{equation}
where $g(z)$ and $h(z)$ are analytic around $\zeta$. Then, as $n\to\infty$,
\begin{equation}
f_n=\frac{h(\zeta)}{\Gamma(-\alpha)}n^{-1-\alpha}\zeta^{-n}\left[1+
O\left(n^{-1}\right)\right].
\end{equation}
\end{theorem}
Applied to $T(z)$, Darboux's theorem implies $t_n=4^n/\sqrt{\pi n^3}+
O\left(n^{-5/2}\right)$ as $n\to\infty$.

\section{Results}

\subsection{Counting secondary structures in RNA}
\label{sec:RNA}

\begin{figure}
\includegraphics[width=80mm]{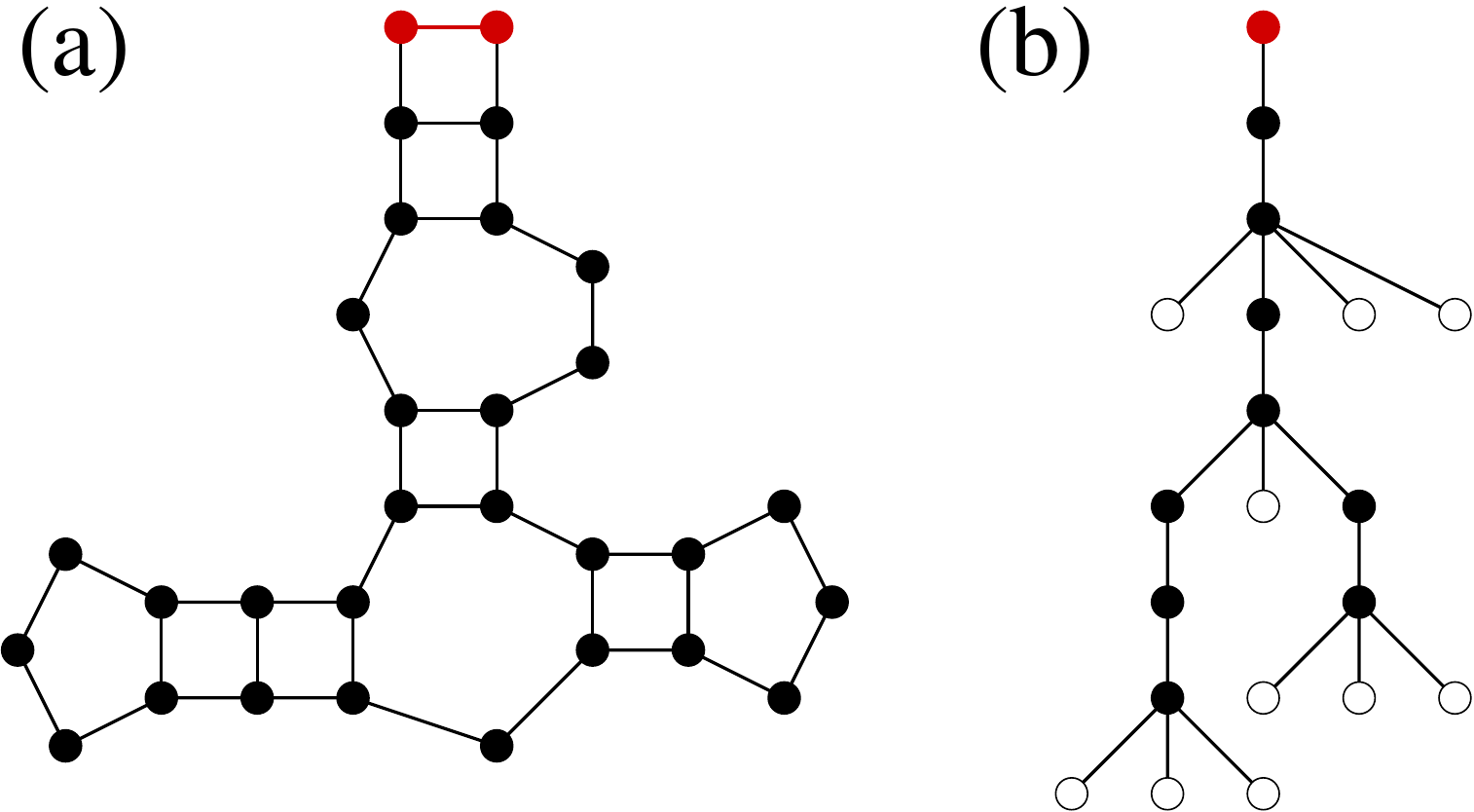}
\caption{\textbf{Tree representation of the secondary structure of RNA
sequences.} (a) Secondary structure of an RNA sequence that starts with a stem.
Stems cannot contain less that two pairs of bases, and hairpin loops cannot be
made of less than three bases. (b) Tree representation of the structure in (a).
Filled circles represent paired bases; empty circles stand for unpaired bases.
For the sake of clarity, the root of the tree in (b) and the corresponding base
pair in the secondary structure (a) are colored.}
\label{fig:RNAfold}
\end{figure}

Figure~\ref{fig:RNAfold}(a) illustrates one possible secondary structure for an
RNA molecule $n=30$ bases long. Some bases are complementary and can pair up
forming a hydrogen bond, some others are not and remain unbound. Sequences of
contiguous paired bases form \emph{stems;} unpaired bases form loops of
different kinds (hairpins, bulges, mutiloops, interior loops\dots). A
description of these structures along with an illustration of them can be found
in \cite{hofacker:1998}.

Determining the specific secondary structure of an RNA molecule is a complex
problem that requires not only a careful energetic minimization, but also 
considerations on the environmental conditions and folding 
kinetics, among others~\cite{chen:2008}. However, some folding constraints 
arise as a consequence of local conditions for energetic stability. Among them, 
two are especially important and were taken into account in early calculations
of the number of realistic RNA secondary structures~\cite{schuster:1994}. Here
we use two general assumptions in agreement with those restrictions: (1) no 
stem can contain less than $s$ pairs, and (2) no hairpin loop can contain less 
than $m$ bases. This notwithstanding, the combinatorial calculations we will be
performing here disregard any further energetic constraints, so the estimation
provided by this method is only an upper bound to the true number of feasible
structures ---because some structures are forbidden on energetic grounds. The
same holds for the circular RNA structures that we will compute later.

We will divide our counting problem in two steps. First, we will count those
foldings starting with a stem ---as the one illustrated in
Figure~\ref{fig:RNAfold}(a). Second, we will take into account that a general
folding consists of several of the former ones joined by free chains
---possibly with chains also at the beginning and/or at the end.

A tree representation of the folding turns out to be more suitable for the
symbolic method. In this representation stems appear as chains of filled dots
($\bullet$) and loops are represented as branches containing an empty dot
($\circ$) per unpaired base and a chain of filled dots per stem branching off
the loop (see Figure~\ref{fig:RNAfold}(b)).

Let $\mathcal{B}$ denote the combinatorial class of all trees representing an
RNA secondary structure starting with a stem and subject to the two above
constraints. Then
\begin{equation}
\mathcal{B}=\mathsf{SEQ}_{\ge s}\big[\{\bullet\}\big]\times
\left(\mathsf{SEQ}\big[\{\circ\}+\mathcal{B}\big]-\mathcal{B}-
\mathsf{SEQ}_{<m}[\{\circ\}]\right).
\label{eq:symbolic}
\end{equation}
The first factor $\mathsf{SEQ}_{\ge s}\big[\{\bullet\}\big]$ stands for the
sequence of $\bullet$ from the root of the tree to the first branching point.
This sequence must have at least $s$ $\bullet$, but its length is otherwise
unlimited ---hence the $\mathsf{SEQ}_{\ge s}$ operator. What one can find at
the first branching event is described by the next factor
$\mathsf{SEQ}\big[\{\circ\}+\mathcal{B}\big]-\mathcal{B}-
\mathsf{SEQ}_{<m}[\{\circ\}]$. The first \textsf{SEQ}
operator means that the number of branches is arbitrary and each branch can
either be a $\circ$ or another tree from the class $\mathcal{B}$ ---hence the
argument $\{\circ\}+\mathcal{B}$. Finally, the term $-\mathcal{B}-
\mathsf{SEQ}_{<m} [\{\circ\}]$ excludes branchings that are not allowed: there
can be neither a single $\mathcal{B}$ branch ---that would mean extending the
previous stem--- nor less than $m$ $\circ$ and nothing else ---that would mean
a hairpin loop with less than $m$ unpaired bases.

Let $B(z)=\sum_{n=0}^{\infty}b_nz^n$ be the generating function of $b_n$, the
number of different $n$-long secondary structures starting with a stem. Since
every $\circ$ (unbounded base) in \eqref{eq:symbolic} contributes $z$ to $B(z)$
and every $\bullet$ (pair of bonded bases) contributes $z^2$ to $B(z)$, we can
translate \eqref{eq:symbolic} as
\begin{equation}
B(z)=\frac{z^{2s}}{1-z^2}\left(\frac{1}{1-z-B(z)}-B(z)-T_m(z)\right),
\label{eq:Bz}
\end{equation}
where $T_m(z)=1+z+\cdots+z^{m-1}$ is the generating function of
$\mathsf{SEQ}_{<m}[\{\circ\}]$.

Once we have characterized the class $\mathcal{B}$, the class of possible RNA
foldings $\mathcal{R}$ can be constructed as
\begin{equation}
\mathcal{R}=\mathsf{SEQ}\big[\{\circ\}+\mathcal{B}\big],
\label{eq:symbolicR}
\end{equation}
i.e., a sequence of arbitrary length (including $n=0$) each of whose components
is either an unpaired base ($\circ$) or a folded structure from $\mathcal{B}$.
In terms of generating functions,
\begin{equation}
R(z)=\frac{1}{1-z-B(z)},
\label{eq:RB}
\end{equation}
where $R(z)=\sum_{n=0}^{\infty}r_nz^n$, $r_n$ being the number of different
$n$-long RNA secondary structures. Eliminating $B(z)$ in this equation and
substituting into \eqref{eq:Bz} leads to the quadratic equation
\begin{equation}
z^{2s}R(z)^2-\big[(1-z)(1-z^2+z^{2s})+z^{2s}T_m(z)\big]R(z)
+1-z^2+z^{2s}=0,
\label{eq:2ndorder}
\end{equation}
whose solution is
\begin{align}
R(z)= &\frac{(1-z)(1-z^2+z^{2s})+z^{2s}T_m(z)-\Delta(z)^{1/2}}{2z^{2s}}, \\
\Delta(z):= &\big[(1-z)(1-z^2+z^{2s})+z^{2s}T_m(z)\big]^2 \nonumber \\
&-4z^{2s}(1-z^2+z^{2s}).
\end{align}
This is Eq.~(43) of Ref.~\cite{hofacker:1998} (beware of a missing factor $2$ in
the left-hand side of that equation).

Suppose $z_*$ is the (single) root of $\Delta(z)$ with the smallest absolute
value. Then $\Delta(z)=(z_*-z)Q(z)$ and the singular part of $R(z)$ will have
the form
\begin{equation}
-\frac{[z_*Q(z)]^{1/2}}{2z^{2s}}\left(1-\frac{z}{z_*}\right)^{1/2}.
\end{equation}
Thus, applying Darboux's theorem we can conclude
\begin{equation}
r_n=\frac{C_s}{2\sqrt{\pi n^3}}z_*^{-n}\left[1+O\left(\frac{1}{n}\right)\right],
\quad C_s:=\frac{Q(z_*)^{1/2}}{2z_*^{2s-1/2}}.
\label{eq:rnasymp}
\end{equation}
For $s=2$, $m=3$ we obtain $z_*=0.540857\dots$ and $C_2=5.263602\dots$, leading
to the well-known result \cite[Table~1]{hofacker:1998} $r_n\sim
1.48483n^{-3/2}(1.84892)^n$.

\subsection{Asymptotic distribution of the number of base pairs}
\label{sec:basepairs}

Now we aim to obtain the asymptotic behavior, when $n,l\to\infty$, of the
distribution $p_{n,l}:=r_{n,l}/r_n$, where $r_{n,l}$ counts the number of RNA
secondary structures having exactly $l$ base pairs. The symbolic method is
easily adapted to obtain $p_{n,l}$. To this end we need to introduce the 
bivariate generating functions
\begin{equation}
R(z,w)=\sum_{n=0}^{\infty}\sum_{l=0}^{\infty}r_{n,l}z^nw^l, \quad
B(z,w)=\sum_{n=0}^{\infty}\sum_{l=0}^{\infty}b_{n,l}z^nw^l,
\end{equation}
where $b_{n,l}$ counts only secondary structures starting with a stem.

Equations \eqref{eq:symbolic} and \eqref{eq:symbolicR} remain valid, but now
every $\circ$ contributes $z$ whereas every $\bullet$ contributes $z^2w$ to
both generating functions (a $\bullet$ is both two bases and a base pair).
Thus, Eqs.~\eqref{eq:Bz} and \eqref{eq:RB} become
\begin{align}
B(z,w) &=\frac{z^{2s}w^s}{1-z^2w}\left(\frac{1}{1-z-B(z,w)}-B(z,w)
-T_m(z)\right), \label{eq:Bzw} \\
R(z,w) &=\frac{1}{1-z-B(z,w)},
\label{eq:RBw}
\end{align}
and we obtain the modified quadratic equation for $R(z,w)$
\begin{equation}
\begin{split}
z^{2s} &w^sR(z,w)^2-\big[(1-z)(1-z^2w+z^{2s}w^s) \\
&+z^{2s}w^sT_m(z)\big]R(z,w)+1-z^2w+z^{2s}w^s=0.
\end{split}
\label{eq:2ndorderw}
\end{equation}

We can interpret $R(z,w)$ as the generating function of the sequence of
polynomials
\begin{equation}
r_n(w):=\sum_{l=0}^{\infty}r_{n,l}w^l
\end{equation}
(notice that $r_{n,l}=0$ if $l>n/2$) and repeat the arguments of the previous
section. Thus, if $z_*(w)$ is the root with smallest absolute value of
\begin{equation}
\begin{split}
\Delta(z,w):= &\big[(1-z)(1-z^2w+z^{2s}w^s)+z^{2s}w^sT_m(z)\big]^2 \\
&-4z^{2s}w^s(1-z^2w+z^{2s}w^s)
\end{split}
\label{eq:Deltazw}
\end{equation}
and $\Delta(z,w)=\big(z_*(w)-z\big)Q(z,w)$, then the singular part of $R(z,w)$
will be
\begin{equation}
-\frac{1}{2z^{2s}w^s}(z_*(w)-z)^{1/2}Q(z,w)^{1/2},
\label{eq:singzw}
\end{equation}
so Darboux's theorem implies (when $n\to\infty$)
\begin{equation}
\begin{split}
r_n(w) &=\frac{C_s(w)}{2\sqrt{\pi n^3}}z_*(w)^{-n}\left[1+O\left(\frac{1}{n}
\right)\right], \\
C_s(w) &:=\frac{Q\big(z_*(w),w\big)^{1/2}}{2z_*(w)^{2s-1/2}w^s}.
\end{split}
\label{eq:rnwasymp}
\end{equation}

Using this information we can obtain the characteristic function of the
probability distribution $p_{n,l}$, for a given $n$, as
\begin{equation}
\phi_n(q):=\sum_{l=0}^{\infty}p_{n,l}e^{iql}=
\frac{r_n\left(e^{iq}\right)}{r_n(1)},
\end{equation}
which, according to eq.~\eqref{eq:rnwasymp}, will behave, asymptotically in
$n$, as
\begin{equation}
\phi_n(q)=A_s\left(e^{iq}\right)\left(\frac{z_*(1)}
{z_*\left(e^{iq}\right)}\right)^{n+2s-\frac{1}{2}}
\left[1+O\left(\frac{1}{n}\right)\right],
\label{eq:characteristic}
\end{equation}
where
\begin{equation}
A_s(w):=\frac{1}{w^s}\left(\frac{Q\big(z_*(w),w\big)}{Q\big(z_*(1),1\big)}
\right)^{1/2}.
\end{equation}
The values of $r_n(1)$, $z_*(1)$, and $Q\big(z_*(1),1\big)$ are those obtained
in Section~\ref{sec:RNA}.

From \eqref{eq:characteristic} it follows
\begin{equation}
\begin{split}
\log\phi_n(q) =&\left(n+2s-\frac{1}{2}\right)
\log\left(\frac{z_*(1)}{z_*\left(e^{iq}\right)}\right) \\
&+\log A_s\left(e^{iq}\right)+O\left(\frac{1}{n}\right) \\
=&\mu_niq-\frac{\sigma_n^2}{2}q^2+O(q^3).
\end{split}
\end{equation}
In other words, the distribution $p_{n,l}$ behaves, as $n\to\infty$, as a
normal distribution in $l$ with mean $\mu_n=\mu n+\mu_0+O\left(n^{-1}\right)$
and standard deviation $\sigma_n=\sigma n^{1/2}+\sigma_0 n^{-1/2}
+O\left(n^{-3/2}\right)$. The precise values depend on $s$ and $m$. For $s=2$,
$m=3$ we obtain $\mu\approx 0.286472\dots$, $\mu_0\approx -0.792076\dots$,
$\sigma\approx 0.255103\dots$, and $\sigma_0\approx 0.247963\dots$ Accordingly,
the number of different phenotypes of a sequence of length $n$ with $l$ paired
bases is given, in the limit $n,l\to\infty$, by
\begin{equation}
r_{n,l}\sim\frac{r_n}{\sqrt{2\pi}\sigma_n}e^{-(l-\mu_n)^2/2\sigma_n^2},
\label{eq:distpairs}
\end{equation}
with $r_n$ as in \eqref{eq:rnasymp}. Equivalent results were obtained 
in~\cite{reidys:2011} and~\cite{poznanovic:2014}.

\subsection{Counting more than one structural element}
\label{sec:2elements}

In this section we are going to count the number of secondary structures with
fixed numbers of base pairs and hairpins. Hairpins are going to be counted
with a variable $u$ ---each hairpin will contribute $u$ to the generating
function. Hairpins are elements of $\mathsf{SEQ}_{\ge m}[\{\circ\}]$, so we
have to separate them out in \eqref{eq:symbolic} and reintroduce them with a
mark $u$. In other words, we need to replace $\mathsf{SEQ}_{<m}[\{\circ\}]$ by
$\mathsf{SEQ}[\{\circ\}]-u\mathsf{SEQ}_{\ge m}[\{\circ\}]$. Since the former
gives rise to the term $T_m(z)$ in \eqref{eq:Bzw}, this operation amounts to
replacing $T_m(z)$ by
\begin{equation}
T_m(z,u)=\frac{1-uz^m}{1-z}
\label{eq:Tmzu}
\end{equation}
in this and subsequent equations.

Now, interpreting $R(z,w,u)$ as the generating function of the bivariate
polynomials
\begin{equation}
r_n(w,u):=\sum_{l=0}^{\infty}\sum_{k=0}^{\infty}r_{n,l,k}w^lu^k,
\end{equation}
$r_{n,l,k}$ being the number of RNA secondary structures with $l$ base pairs
and $k$ hairpins, we can obtain the asymptotic behavior of the probability
distribution $p_{n,l,k}:=r_{n,l,k}/r_n$ through that of its characteristic
function
\begin{equation}
\phi_n(\vec{q})=\frac{r_n\left(e^{iq_p},e^{iq_h}\right)}{r_n(1,1)}, \qquad
\vec{q}:=(q_p,q_h).
\end{equation}
Following the procedure explained in the previous section we find
\begin{equation}
\begin{split}
\log\phi_n(\vec{q})=&
\left(n+2s-\frac{1}{2}\right)\log\left(\frac{z_*(1,1)}{z_*\left(e^{iq_p},
e^{iq_h}\right)}\right) \\
&+\log A_s\left(e^{iq_p},e^{iq_h}\right)+O\left(\frac{1}{n}\right),
\end{split}
\label{eq:logphi}
\end{equation}
with
\begin{equation}
A_s(w,u):=\frac{1}{w^s}\left(\frac{Q\big(z_*(w,u),w,u\big)}{Q\big(z_*(1,1),
1,1\big)}\right)^{1/2},
\end{equation}
$z_*(w,u)$ being the singularity of $R(z,w,u)$ with smallest absolute value,
and $Q(z,w,u)$ defined as in \eqref{eq:Deltazw}, \eqref{eq:singzw}, with
$T_m(z)$ replaced by $T_m(z,u)$ defined in Eq.~\eqref{eq:Tmzu}. If we now
identify
\begin{equation}
\log\phi_n(\vec{q})=\mu^p_niq_p+\mu^h_niq_h-\frac{1}{2}\vec{q}\cdot
\tens{\Sigma}_n\cdot\vec{q}^{\mathsf{T}}+O\left(\|\vec{q}\|^3\right),
\label{eq:logcharbiv}
\end{equation}
we obtain the mean vector $(\mu^p_n,\mu^h_n)$ and covariance matrix
$\tens{\Sigma}_n$ of a bivariate normal distribution. For instance, setting
$s=2$, $m=3$ we get
\begin{equation}
\begin{split}
&\mu_n^p= (0.286472\dots) n-(0.792076\dots)+O\left(n^{-1}\right), \\
&\mu_n^h= (0.0378631\dots) n+(0.308604\dots)+O\left(n^{-1}\right), \\
&\tens{\Sigma}_n^{pp}= (0.0650779\dots) n+(0.126513\dots)
+O\left(n^{-1}\right), \\
&\tens{\Sigma}_n^{hh}= (0.0115908\dots) n+(0.0164609\dots)
+O\left(n^{-1}\right), \\
&\tens{\Sigma}_n^{ph}= (-0.00274347\dots) n+(0.00918949\dots)
+O\left(n^{-1}\right).
\end{split}
\end{equation}

Thus, asymptotically,
\begin{equation}
\begin{split}
r_{n,l,k}\sim &\frac{r_n}{2\pi|\tens{\Sigma}_n|^{1/2}} \\
&\times\exp\left\{
-\frac{1}{2}(l-\mu^p_n,k-\mu^h_n)\cdot\tens{\Sigma}_n^{-1}\cdot
(l-\mu^p_n,k-\mu^h_n)^{\mathsf{T}}\right\}.
\end{split}
\label{eq:asymdist}
\end{equation}

Obtaining the marginal distribution of base pairs amounts to setting $q_h=0$ in
\eqref{eq:logcharbiv}. One can easily check that it correspond to the
distribution \eqref{eq:distpairs}. Likewise, the marginal distribution of
hairpins follows from setting $q_p=0$ in \eqref{eq:logcharbiv}. It turns out to
be a normal distribution with mean $\mu_n^h$ and variance
$\tens{\Sigma}_n^{hh}$.

New structural elements can be counted in a similar vein, and their
corresponding asymptotic distribution will be multivariate normal distributions
whose parameters can be determined as we have done in this section. Analogous
results for multivariate distributions of structural motifs can be found 
in~\cite{poznanovic:2014}.

\subsection{Counting secondary structures of circular RNAs}
\label{sec:rings}

\begin{figure}
\includegraphics[width=70mm]{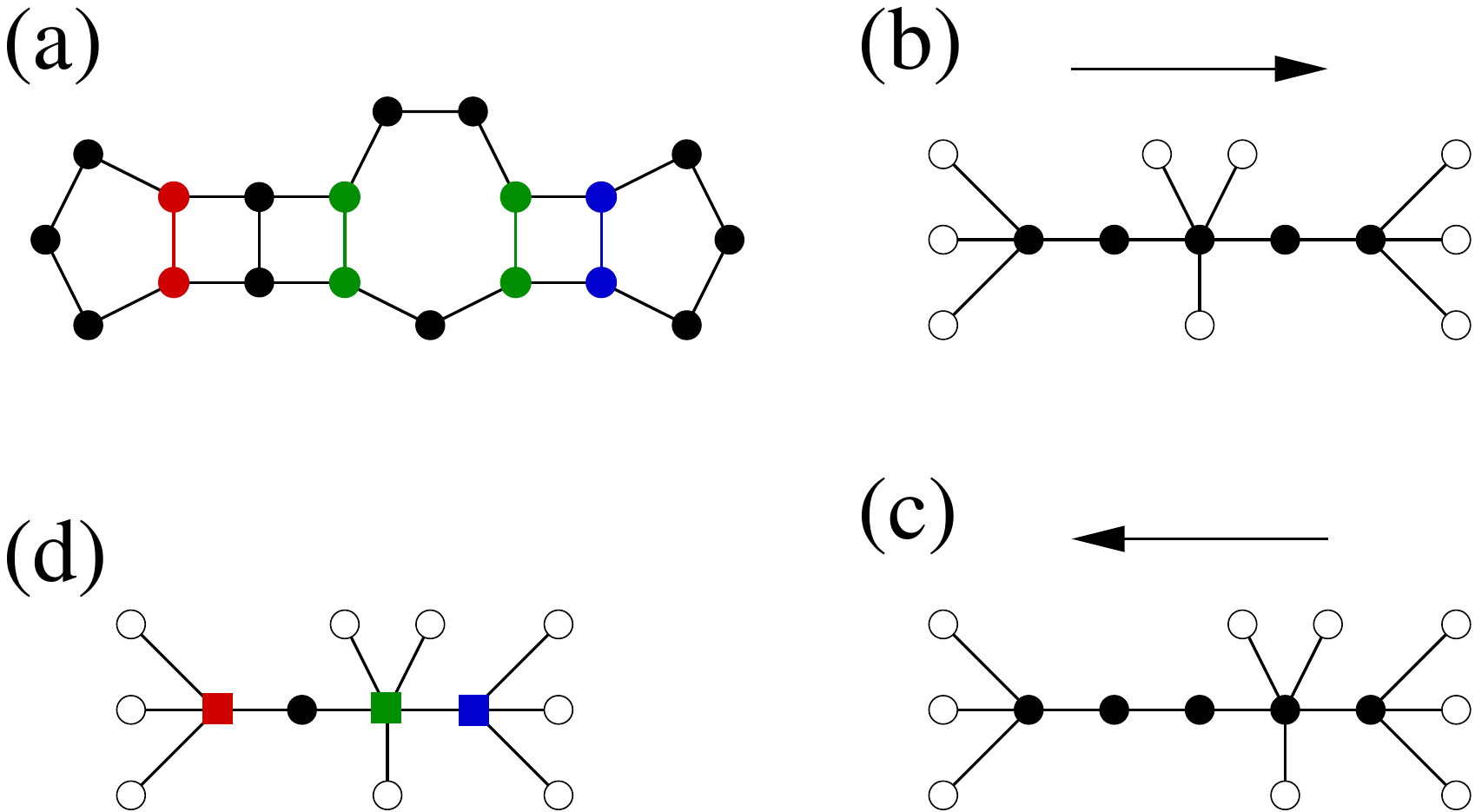}
\caption{\textbf{Tree representation of the secondary structure of circular
RNAs.} (a) Secondary structure of a circular RNA sequence. (b) Tree
representation of the structure in (a) as read starting from the leftmost
hairpin. (c) Tree representation of the same structure but read from the
rightmost hairpin. (d) New tree representation in which square nodes mark the
extremes of the stems ---hence leaves (empty circles) hang from these nodes.
Each square counts one base pair for each stem meeting at it. (Colors are
meant to help understand the association between base pairs and square nodes.)
Notice that this tree is uniquely defined by the RNA structure regardless of
the way we read it.}
\label{fig:RNAring}
\end{figure}

Let now $\mathcal{V}$ denote the combinatorial class containing all secondary
structures of circular RNAs. As for open sequences, counting is better done 
using the tree representation of Fig.~\ref{fig:RNAfold}. If secondary 
structures of linear sequences are encoded in rooted trees, those of circular 
sequences, for which any base pair can act as a root, would correspond to
unrooted trees. There is an ambiguity though when transforming the rooted tree
representation into an unrooted one. The rules to transform structures into
trees are directional, as illustrated in Fig.~\ref{fig:RNAring}. To avoid that
we introduce a new type of node, a square, to mark the extremes of all stems
meeting at a hairpin, a multiloop, or a bulge. The square is understood to
represent a base pair for each stem meeting at it. With this new representation
each secondary structure of a circular RNA uniquely determines a tree with two
types of inner nodes ---filled circles and squares--- and empty circles for
leaves, regardless of the direction we choose to read the structure. 

We will need a new combinatorial class to obtain $\mathcal{V}$, namely
\begin{equation}
\mathcal{B}_k=\mathsf{SEQ}_k\big[\{\bullet\}\big]\times
\left(\mathsf{SEQ}\big[\{\circ\}+\mathcal{B}\big]-\mathcal{B}-
\mathsf{SEQ}_{<m}[\{\circ\}]\right),
\label{eq:Bk}
\end{equation}
the class of secondary RNA structures starting with a stem of exactly $k$ base
pairs. Notice that \eqref{eq:symbolic} implies that $\mathcal{B}=\sum_{k\ge s}
\mathcal{B}_k$, and it follows from \eqref{eq:symbolic} and \eqref{eq:Bk} that
\begin{equation}
B_k(z)=z^{2k-2s}(1-z^2)B(z).
\end{equation}

Counting unrooted trees is a more complicated issue than counting rooted trees.
As a matter of fact, the strategy to do it is to reduce the problem to counting
rooted trees. This is achieved thanks to a so-called \emph{dissymetry theorem}
that relates both classes of trees \cite[\S 4.1]{bergeron:1998}. If
$\mathcal{F}$ denotes a class of rooted trees and $\mathcal{G}$ denotes that
of their corresponding unrooted trees, then
\begin{equation}
\mathcal{G}^{\bullet}+\mathcal{G}^{\bullet\!\!-\!\!\bullet}=
\mathcal{G}+\mathcal{F}\times\mathcal{F},
\end{equation}
where $\mathcal{G}^{\bullet}$ denotes the class of unrooted trees with a marked
node, and $\mathcal{G}^{\bullet\!\!-\!\!\bullet}$ denotes the class of unrooted
trees with a marked link. In our case, $\mathcal{G}$ stands for $\mathcal{V}$,
the class we want to count. As for $\mathcal{F}\times\mathcal{F}$, an analysis
of the proof of the theorem reveals that the $\mathcal{F}$s involved arise as a
result of removing links in trees of $\mathcal{G}$. Thus, for the kind of trees
we aim at counting we need to adapt this result, because links in $\mathcal{V}$
are part of a stem, and stems must have at least $s$ base pairs. Also, as
leaves (empty circles) are never the root of a tree, the argument can focus on
inner nodes and inner links.

Consider $v\in\mathcal{V}$. Removing an inner link in $v$ yields two trees, one
belonging to $\mathcal{B}_j$ and another one belonging to $\mathcal{B}_k$, such
that $j,k\ge 1$ and $j+k\ge s$. Therefore
\begin{equation}
\mathcal{F}_s:=\mathcal{F}\times\mathcal{F}=\sum_{j+k\ge s \atop j,k\ge 1}
\mathcal{B}_j\times\mathcal{B}_k.
\end{equation}
Let us now mark a link of $v$ to transform it into an element of
$\mathcal{V}^{\bullet\!\!-\!\!\bullet}$. Two rooted trees from $\mathcal{B}_j$
and $\mathcal{B}_k$ ---with the same index constraints--- hang from both sides
of the marked link. Since the order of these two trees is irrelevant,
\begin{equation}
\mathcal{V}^{\bullet\!\!-\!\!\bullet}=\frac{1}{2}(\mathcal{F}_s+
\mathcal{D}_s), \qquad \mathcal{D}_s:=\sum_{2j\ge s}
\mathsf{DIAG}(\mathcal{B}_j),
\end{equation}
using the idea behind the definition of $\mathsf{CYC}_2$
(Sec.~\ref{sec:brief}). Finally, if we mark a $\bullet$ node as root, the two
hanging branches are one tree from $\mathcal{B}_j$ and another one from
$\mathcal{B}_k$, such that $j,k\ge 1$ and $j+k\ge s-1$; but if we mark a
{\scriptsize$\blacksquare$} node as root, the resulting tree is formed by
a ring from which either leaves ($\circ$) or $\mathcal{B}$ trees hang. Thus
\begin{equation}
\begin{split}
\mathcal{V}^{\bullet}=& \{\bullet\}\times\frac{1}{2}(\mathcal{F}_{s-1}+
\mathcal{D}_{s-1})+\mathsf{CYC}\big[\{\circ\}+\mathcal{B}\big] \\
&-\mathcal{B}\times
\mathsf{SEQ}_{<m}\big[\{\circ\}\big]-\mathsf{CYC}_2[\mathcal{B}],
\end{split}
\end{equation}
where the two last terms stand for the removal of hairpins not allowed by the
constraints ($\mathcal{B}\times\mathsf{SEQ}_{<m}\big[\{\circ\}\big]$) and of
cycles containing just two $\mathcal{B}$ trees and no $\circ$ leave
($\mathsf{CYC}_2[\mathcal{B}]$) ---which would be indistinguishable from longer
stems. Summarizing,
\begin{equation}
\begin{split}
\mathcal{V}=& \frac{1}{2}\big(\{\bullet\}\times\mathcal{F}_{s-1}-
\mathcal{F}_s+\{\bullet\}\times\mathcal{D}_{s-1}+\mathcal{D}_s\big) \\
&+\mathsf{CYC}\big[\{\circ\}+\mathcal{B}\big]-\mathcal{B}\times
\mathsf{SEQ}_{<m}\big[\{\circ\}\big]-\mathsf{CYC}_2[\mathcal{B}].
\end{split}
\end{equation}

Now,
\begin{equation}
\begin{split}
F_s(z) &=\frac{B(z)^2\big(1-z^2\big)^2}{z^{4s}}\sum_{j+k\ge s \atop j,k\ge 1}
z^{2(j+k)} \\
&=\frac{B(z)^2\big(1-z^2\big)^2}{z^{4s}}\sum_{l=s}^{\infty}(l-1)z^{2l} \\
&=\frac{B(z)^2}{z^{2s}}\left[s-1-(s-2)z^2\right],
\end{split}
\end{equation}
and similarly
\begin{equation}
\begin{split}
F_{s-1}(z) &=\frac{B(z)^2\big(1-z^2\big)^2}{z^{4s}}\sum_{l=s-1}^{\infty}
(l-1)z^{2l} \\
&=\frac{B(z)^2}{z^{2s+2}}\left[s-2-(s-3)z^2\right],
\end{split}
\end{equation}
so the generating function of $\{\bullet\}\times\mathcal{F}_{s-1}-
\mathcal{F}_s$ is
\begin{equation}
\begin{split}
\frac{B(z)^2}{z^{2s}} &\left[s-2-(s-3)z^2\right]-
\frac{B(z)^2}{z^{2s}}\left[s-1-(s-2)z^2\right] \\
&=-\frac{B(z)^2}{z^{2s}}\big(1-z^2\big).
\end{split}
\end{equation}
On the other hand,
\begin{equation}
D_s(z)=\sum_{2k\ge s}B_k\big(z^2\big)
=\frac{B\big(z^2\big)\big(1-z^4\big)}{z^{4s}}\sum_{2k\ge s}z^{4k}
\end{equation}
and
\begin{equation}
D_{s-1}(z)=\frac{B\big(z^2\big)\big(1-z^4\big)}{z^{4s}}
\sum_{2k+1\ge s}z^{4k},
\end{equation}
so the generating function of $\{\bullet\}\times\mathcal{D}_{s-1}+
\mathcal{D}_s$ is
\begin{equation}
\begin{split}
\frac{B\big(z^2\big)\big(1-z^4\big)}{z^{4s}} &\left(\sum_{2k+1\ge s}
z^{2(2k+1)}+\sum_{2k\ge s}z^{2(2k)}\right) \\
&=\frac{B\big(z^2\big)\big(1-z^4\big)}{z^{4s}}\sum_{l=s}^{\infty}z^{2l}
=\frac{B\big(z^2\big)\big(1+z^2\big)}{z^{2s}}.
\end{split}
\end{equation}
If we take into account that the generating function of
$\mathsf{CYC}_2[\mathcal{B}]$ is
\begin{equation}
\frac{1}{2}\left[B(z)^2+B\big(z^2\big)\right],
\end{equation}
we finally obtain
\begin{equation}
\begin{split}
V(z)=& \frac{1}{2z^{2s}}\Big[B\big(z^2\big)\big(1+z^2-z^{2s}\big)-
B(z)^2\big(1-z^2+z^{2s}\big)\Big] \\
&-\sum_{k=1}^{\infty}\frac{\varphi(k)}{k}\log\left[1-z^k-B\big(z^k\big)
\right]-B(z)T_m(z),
\end{split}
\end{equation}
or using \eqref{eq:RB},
\begin{equation}
\begin{split}
V(z)=& \frac{1}{2z^{2s}}\Big[B\big(z^2\big)\big(1+z^2-z^{2s}\big)-
B(z)^2\big(1-z^2+z^{2s}\big)\Big] \\
&+\sum_{k=1}^{\infty}\frac{\varphi(k)}{k}\log R\big(z^k\big)-B(z)T_m(z).
\end{split}
\label{eq:Vz}
\end{equation}
Incidentally, $B(z)$ is derived straight away from \eqref{eq:RB} as
\begin{equation}
B(z)=\frac{(1-z)(1-z^2+z^{2s})-z^{2s}T_m(z)-\Delta(z)^{1/2}}{2(1-z^2+z^{2s})}.
\end{equation}

\begin{table}
\caption{Number of secondary structures ---excluding the unfolded chain---
of a circular RNA sequence of length $n$ (we have set $s=2$ and $m=3$).}
\label{tab:1}
\begin{center}
\begin{tabular}{cccccc}
\hline\noalign{\smallskip}
\multicolumn{1}{c}{$n$} & \# struct. &
\multicolumn{1}{c}{$n$} & \# struct. &
\multicolumn{1}{c}{$n$} & \# struct. \\
\noalign{\smallskip}\hline\noalign{\smallskip}
10 &  1 & 20 &   105 & 30 &   20423 \\
11 &  1 & 21 &   166 & 31 &   35091 \\
12 &  3 & 22 &   287 & 32 &   60838 \\
13 &  3 & 23 &   486 & 33 &  105169 \\
14 &  6 & 24 &   816 & 34 &  182728 \\
15 &  7 & 25 &  1364 & 35 &  317068 \\
16 & 14 & 26 &  2368 & 36 &  552059 \\
17 & 20 & 27 &  4011 & 37 &  961008 \\
18 & 38 & 28 &  6972 & 38 & 1677222 \\
19 & 59 & 29 & 11811 & 39 & 2928607 \\
\noalign{\smallskip}\hline
\end{tabular}
\end{center}
\end{table}

Table~\ref{tab:1} lists the coefficients of $V(z)$ up to $n=39$ ---discounting
1 for the unfolded chain. For long chains we can obtain an asymptotic formula
out of \eqref{eq:Vz}. Despite its appearance ---especially because of the
presence of an infinite series---, finding the singularity $z_*$ closest to the
origin of $V(z)$ is an easy task. That singularity is to be found in the
functions $B(z)$ and $R(z)$, as a root of $\Delta(z)$. We know $0<z_*<1$
because all coefficients in the power series $V(z)$ are larger than $1$ (as a
matter of fact, for $s=2$, $m=3$ we already found $z_*=0.540857\dots$). This
means that the corresponding root of terms of the form $\Delta\big(z^k\big)$,
with $k>1$, will be $z_*^{1/k}>z_*$. In other words, all terms $B\big(z^2\big)$
and $R\big(z^k\big)$ with $k>1$ are analytic at $z_*$. The only possibly
competing singularity would come from  a root of $R(z)$ in $\log R(z)$.  But
$R(z)=0$ implies $1-z^2+z^{2s}=0$, whose solutions for $s=2$ are $\pm e^{\pm
i\pi/6}$ and therefore their modulus is larger than $z_*$.

From this discussion we conclude that the singular terms of $V(z)$ that will
contribute to the asymptotic behavior of its coefficients are those containing
$B(z)$, $B(z)^2$ and $\log R(z)$. Accordingly, $V(z)$ can be written, when
$\Delta(z)\to 0$, as
\begin{align*}
V(z)=&\, \zeta(z)+\frac{\big[(1-z)(1-z^2+z^{2s})-z^{2s}T_m(z)\big]
\Delta(z)^{1/2}}{4z^{2s}(1-z^2+z^{2s})} \\
&+\frac{T_m(z)\Delta(z)^{1/2}}{2(1-z^2+z^{2s})}
-\frac{\Delta(z)^{1/2}}{(1-z)(1-z^2+z^{2s})+z^{2s}T_m(z)} \\
&-\frac{\Delta(z)^{3/2}}{3\big[(1-z)(1-z^2+z^{2s})+z^{2s}T_m(z)\big]^3}
+O\left(\Delta(z)^{5/2}\right) \\
=&\, \zeta(z) \\
&+\frac{\Delta(z)^{3/2}}{4z^{2s}(1-z^2+z^{2s})
\big[(1-z)(1-z^2+z^{2s})+z^{2s}T_m(z)\big]} \\
&-\frac{\Delta(z)^{3/2}}{3\big[(1-z)(1-z^2+z^{2s})+z^{2s}T_m(z)\big]^3}
+O\left(\Delta(z)^{5/2}\right),
\end{align*}
where $\zeta(z)$ is an analytic function in a circle containing $z_*$. Now,
since $(1-z)(1-z^2+z^{2s})+z^{2s}T_m(z)=2z^s(1-z^2+z^{2s})^{1/2}+
O\big(\Delta(z)\big)$ follows from the very definition of $\Delta(z)$, the
expression above simplifies to
\begin{equation}
V(z)=\zeta(z)+\frac{\Delta(z)^{3/2}}{12z^{3s}(1-z^2+z^{2s})^{3/2}}
+O\left(\Delta(z)^{5/2}\right).
\end{equation}

As in Sec.~\ref{sec:RNA} we can write $\Delta(z)=(z_*-z)Q(z)$, so near $z_*$
\begin{equation}
\begin{split}
V(z)= &\zeta(z_*)+\frac{Q(z_*)^{3/2}}{12z_*^{3s-\frac{3}{2}}
(1-z_*^2+z_*^{2s})^{3/2}}\left(1-\frac{z}{z_*}\right)^{3/2} \\
&+O\left(\left(1-\frac{z}{z_*}\right)^{5/2}\right),
\end{split}
\end{equation}
and then Darboux's theorem yields
\begin{equation}
\begin{split}
v_n &=\frac{3K_s}{4\sqrt{\pi n^5}}z_*^{-n}\left[1+
O\left(\frac{1}{n}\right)\right], \\
K_s &:=\frac{Q(z_*)^{3/2}}{12z_*^{3s-\frac{3}{2}}
(1-z_*^2+z_*^{2s})^{3/2}}.
\end{split}
\end{equation}
For $s=2$, $m=3$ we obtain $K_2=3.445906\dots$, so we find the asymptotic
estimate for the number of structures of circular RNA sequences $v_n\sim
1.45811n^{-5/2}(1.84892)^n$.

\subsection{Base pairs and hairpins in circular RNAs}
\label{sec:elementsRings}

We can introduce $v_{n,l,k}$, the number of circular RNAs with $l$ base pairs and
$k$ hairpins, and $V(z,w,u)$, the generating function of the bivariate
polynomials
\begin{equation}
v_n(w,u)=\sum_{l=0}^{\infty}\sum_{k=0}^{\infty}v_{n,l,k}w^lu^k.
\end{equation}
This generating function can be obtained, following the steps in
sections~\ref{sec:basepairs} and \ref{sec:2elements}, to be
\begin{equation}
\begin{split}
V(z,w,u)=& \frac{1}{2z^{2s}w^s}\Big[B\big(z^2,w^2,u^2\big)
\big(1+z^2w-z^{2s}w^s\big) \\
&-B(z,w,u)^2\big(1-z^2w+z^{2s}w^s\big)\Big] \\
&+\sum_{k=1}^{\infty}\frac{\varphi(k)}{k}\log R\big(z^k,w^k,u^k\big) \\
&-B(z,w,u)T_m(z,u).
\end{split}
\label{eq:Vzwu}
\end{equation}
It follows from this equation and the asymptotic analysis in the previous
section that the characteristic function $\phi_n(\vec{q})$ of the probability
distribution $p_{n,l,k}:=v_{n,l,k}/v_n$ is asymptotically given by
\begin{equation}
\begin{split}
\log\phi_n(\vec{q})=&
\left(n+3s-\frac{3}{2}\right)\log\left(\frac{z_*(1,1)}{z_*\left(e^{iq_p},
e^{iq_h}\right)}\right) \\
&+\log D_s\left(e^{iq_p},e^{iq_h}\right)+
O\left(\frac{1}{n}\right),
\end{split}
\end{equation}
where
\begin{equation}
D_s(w,u):=\left[\frac{Q\big(z_*(w,u),w,u\big)\big(1-z_*(1,1)^2+
z_*(1,1)^{2s}\big)}{w^sQ\big(z_*(1,1),1,1\big)\big(1-z_*(w,u)^2+
z_*(w,u)^{2s}\big)}\right]^{3/2}.
\end{equation}
As expected, the leading term is the same as in \eqref{eq:logphi}.

Identifying this expression with the expansion \eqref{eq:logcharbiv} we obtain,
for $s=2$, $m=3$, the probability distribution \eqref{eq:asymdist} with
\begin{equation}
\begin{split}
&\mu_n^p= (0.286472\dots) n+(0.773395\dots)+O\left(n^{-1}\right), \\
&\mu_n^h= (0.0378631\dots) n+(0.681247\dots)+O\left(n^{-1}\right), \\
&\tens{\Sigma}_n^{pp}= (0.0650779\dots) n-(0.060170\dots)
+O\left(n^{-1}\right), \\
&\tens{\Sigma}_n^{hh}= (0.0115908\dots) n-(0.0258221\dots)
+O\left(n^{-1}\right), \\
&\tens{\Sigma}_n^{ph}= (-0.00274347\dots) n+(0.0427301\dots)
+O\left(n^{-1}\right).
\end{split}
\end{equation}

\section{Discussion and conclusions}

The symbolic method can be extended to the case of circular RNAs in order to 
calculate the total number of closed secondary structures for sequences of 
length $n$ and the asymptotic distributions of the number of structures with 
specific moieties. Circularization of RNA eliminates some degrees of freedom 
that translate into a number of secondary structures $n$-fold lower, as 
compared to the open linear counterpart. The exponent $b=5/2$ also appears
in the enumeration of unrooted trees~\cite{flajolet:2009}, of which circular
RNAs are a particular case.  

The relationship between structure and function in circular RNAs has to be 
stronger than in linear RNAs, due at least to the non-coding nature of most of
the former. From an evolutionary viewpoint, circularization of RNAs might be a 
low-cost procedure to seek new molecular functions. Closed structures differ in
essential ways from their open counterparts in their stability properties, and
may as well bind different molecules due, for instance, to the sequences
brought together when open ends are covalently closed~\cite{jeck:2014}. At the
same time, the number of available folds decreases under circularization by
essentially a factor $n$. This severe decrease in structural repertoire with
respect to the open molecule implies that, on average, there are $n$ times more
sequences that fold into a closed structure than into an open structure of the
same length. The mutational robustness of closed structures is therefore very
much enhanced.

The enumeration of circular RNA structures with pseudoknots is an open 
problem with relevance, among others, to better understand the {\it in vivo} 
conformations adopted by viroids~\cite{flores:2012} and other circular RNAs
encoded in genomes, and the identification of their hypothetical interacting 
sites. A combination of the symbolic method and the additional techniques here 
used for circular RNA might facilitate the achievement of that goal.

\section{Acknowledgements}

The authors acknowledge conversations with Christine Heitsch. This work was
supported by the Spanish Ministerio de Econom\'{\i}a y Competitividad and FEDER
funds from the EU (grant numbers FIS2014-57686-P and FIS2015-64349-P).

\section*{References}

\bibliography{evol-coop}

\end{document}